\documentclass[12pt]{article}
\usepackage[usenames,dvipsnames]{color}
\usepackage{graphicx,microtype}
\usepackage{titling}
\usepackage{amsmath}
\usepackage{authblk}
\usepackage{textgreek}
\usepackage{xr-hyper}
\usepackage{setspace}
\usepackage[bookmarks=false,colorlinks]{hyperref}
\hypersetup{linkcolor=RubineRed,citecolor=Plum,filecolor=Plum,urlcolor=Aquamarine}

\usepackage{xspace}
\newcommand{\sto}{SrTiO\textsubscript{3}\xspace}

\newcommand{\lbo}{LiNbO\textsubscript{3}\xspace}

 \newcommand{\simapprox}{\raisebox{0.5ex}{\texttildelow}}
\begin{document}

\title{Observation of polarization density waves in SrTiO$_3$}

\begin{singlespace}

\author[1,2]{Gal Orenstein\footnote{These authors contributed equally}}
\author[1,2,3]{Viktor Krapivin$^*$}
\author[4]{Yijing Huang}
\author[5]{Zhuquan Zhang}
\author[1,2]{Gilberto de la Pe\~{n}a Mu\~{n}oz}
\author[1,2]{Ryan A. Duncan}
\author[1,2]{Quynh Nguyen}
\author[3,6]{Jade Stanton}
\author[6]{Samuel Teitelbaum}
\author[7]{Hasan Yavas}
\author[7]{Takahiro Sato}
\author[7]{Matthias C. Hoffmann}
\author[7]{Patrick Kramer}
\author[8]{Jiahao Zhang}
\author[9]{Andrea Cavalleri}
\author[10]{Riccardo Comin}
\author[11]{Mark P. M. Dean}
\author[12]{Ankit S. Disa}
\author[9]{Michael F\"{o}rst}
\author[13,14]{Steven L. Johnson}
\author[15]{Matteo Mitrano}
\author[8]{Andrew M. Rappe}
\author[1,2,3]{David Reis}
\author[7]{Diling Zhu}
\author[5]{Keith A. Nelson}
\author[1,2]{Mariano Trigo\footnote{mtrigo@slac.stanford.edu}}

\affil[1]{Stanford PULSE Institute, SLAC National Accelerator Laboratory, Menlo Park, California 94025, USA}
\affil[2]{Stanford Institute for Materials and Energy Sciences, SLAC National Accelerator Laboratory, Menlo Park, California 94025, USA}
\affil[3]{Department of Applied Physics, Stanford University, Stanford, California 94305, USA}
\affil[4]{Department of Physics, University of Illinois at Urbana-Champaign, Urbana, Illinois 61820, USA}
\affil[5]{Department of Chemistry, Massachusetts Institute of Technology, Cambridge, Massachusetts 02139, USA}
\affil[6]{Department of Physics, Arizona State University, Tempe, Arizona 85281, USA.}
\affil[7]{Linac Coherent Light Source, SLAC National Accelerator Laboratory, Menlo Park, California 94025, USA}
\affil[8]{Department of Materials Science and Engineering, University of Pennsylvania, Philadelphia 19104, USA}
\affil[9]{Max Planck Institute for the Structure and Dynamics of Matter, Hamburg, Germany}
\affil[10]{Department of Physics, Massachusetts Institute of Technology, Cambridge, Massachusetts 02139, USA}
\affil[11]{Condensed Matter Physics and Materials Science Department, Brookhaven National Laboratory, Upton, New York 11973, USA}
\affil[12]{School of Applied \& Engineering Physics, Cornell University, Ithaca, NY 14853}
\affil[13]{Institute for Quantum Electronics, Physics Department, ETH Zurich, CH-8093 Zurich, Switzerland}
\affil[14]{SwissFEL, Paul Scherrer Institut, CH-5232 Villigen PSI, Switzerland}
\affil[15]{Department of Physics, Harvard University, Cambridge, Massachusetts 02138, USA}

\date{}

\maketitle

\end{singlespace}

{\bf

The nature of the “failed” ferroelectric transition in \sto has been a long-standing puzzle in condensed matter physics \cite{mullerSrTiIntrinsicQuantum1979,rowleyFerroelectricQuantumCriticality2014,aschauerCompetitionCooperationAntiferrodistortive2014,Yamanaka_2000}. A compelling explanation is the competition between ferroelectricity and an instability with a mesoscopic modulation of the polarization\cite{mullerIndicationNovelPhase1991,coakQuantumCritical2020,fauqueMesoscopicFluctuatingDomains2022,guzman-verriLamellarFluctuationsMelt2023}. 
These polarization density waves, which should become especially strong near the quantum critical point\cite{rowleyFerroelectricQuantumCriticality2014,coakQuantumCritical2020}, break local inversion symmetry and are difficult to probe with conventional x-ray scattering methods. 
Here we combine a femtosecond x-ray free electron laser (XFEL) with THz coherent control methods to probe inversion symmetry breaking at finite momenta\cite{trigoFouriertransformInelasticXray2013} and visualize the instability of the polarization on nanometer lengthscales in \sto.
We find polar-acoustic collective modes that are soft particularly at the tens of nanometer lengthscale. These precursor collective excitations provide evidence for the conjectured mesoscopic modulated phase in \sto\cite{guzman-verriLamellarFluctuationsMelt2023,coakQuantumCritical2020}.

}

\sto exhibits typical characteristics of an incipient ferroelectric (FE) material, such as an increase in the dielectric constant \cite{weaverDielectricPropertiesSingle1959} and a polar mode softening \cite{shiraneLatticeDynamicalStudy1101969} upon cooling. However, quantum fluctuations at low temperatures prevent long-range polar order in this material, rendering it a quintessential example of a quantum paraelectric \cite{mullerSrTiIntrinsicQuantum1979}.
Nevertheless, \sto{} exhibits mesoscopic fluctuations of the FE polarization and develops polar nanoregions at low temperatures, which have a significant impact on the properties of the material\cite{Bussmann2009precursor,bussmann2006intrinsic,Laguta2005nmr,Biancoli2015}.
A FE phase in \sto is readily obtained by various methods such as strain \cite{haeniRoomtemperatureFerroelectricityStrained2004,xuStraininducedRoomtemperatureFerroelectricity2020}, calcium substitution \cite{bedborz1984randomness} or oxygen isotope substitution \cite{itoh1999oxygen}. 
More recently, FE features have been induced by driving vibrational modes using ultrafast THz and mid-infrared pulses\cite{liTerahertzFieldInduced2019,novaMetastableFerroelectricityOptically2019,ChengTerahertzDriven2023}. 
These properties of \sto indicate its close proximity to a FE instability at low temperatures.

The FE instability is well-established and is associated with the softening of an optical polar mode at zero wavevector. However, several susceptibility and neutron scattering measurements have led to speculation that fluctuating polar-acoustic modes with finite wavevectors strongly affect the quantum paraelectric regime at low temperature \cite{mullerIndicationNovelPhase1991, vacherAnomaliesTAPhononBranches1992,rowleyFerroelectricQuantumCriticality2014,coakQuantumCritical2020,fauqueMesoscopicFluctuatingDomains2022}.
These results may indicate the existence of another structural instability in \sto that is different from the homogeneous FE state, with strong lattice fluctuations on the nanometer lengthscale, potentially heralding a novel phase with spatially modulated polarization \cite{coakQuantumCritical2020,mullerIndicationNovelPhase1991,guzman-verriLamellarFluctuationsMelt2023}. 
However, the crucial element of this putative polar-acoustic regime-- the mesoscopic polar characteristics of \sto{}-- could not be determined by these measurements and had to be indirectly inferred from models.

\begin{figure*}
\centering
\includegraphics[width=0.9\linewidth]{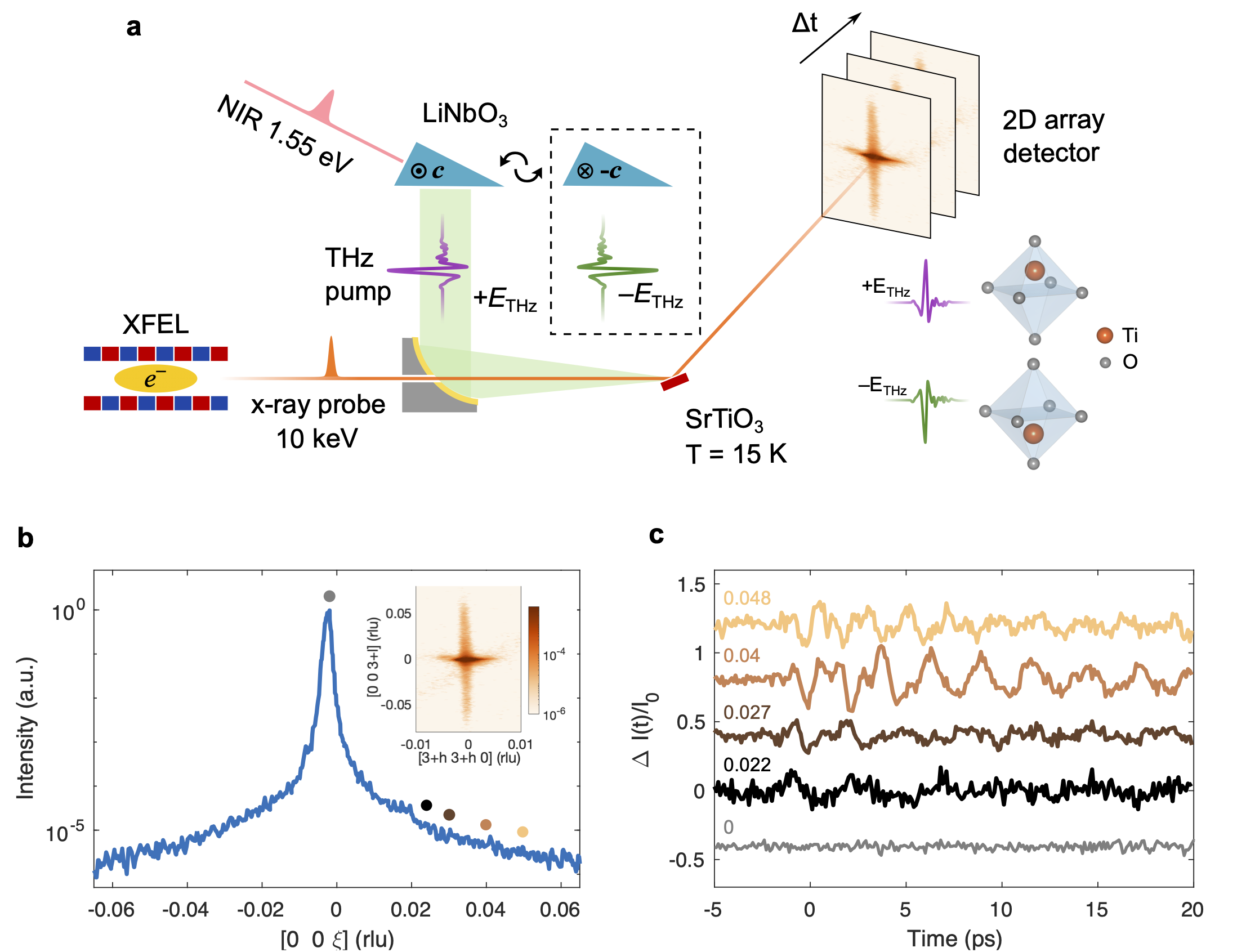}
\caption{\label{fig:1} 
{\bf Ultrafast hard x-ray scattering under strong THz excitation.} (a) Schematic illustration of the experimental setup. Pulses with a spectrum centered at 0.5 THz were produced by optical rectification of 100 fs near-IR pulses centered at 800 nm. 
Two \lbo{} crystals with FE polarizations pointing in opposite directions enable swapping the polarity of the emitted radiation by vertically translating the \lbo{} prisms into the incident near IR beam. The THz pulse is focused on to the \sto{} sample using a parabolic mirror and is colinear with the X-ray probe pulse with photon energy of 10 keV. 
The inset of (a) shows an illustration of the polar displacement of the Ti atom upon excitation with fields of opposite polarity.
(b) Integrated intensity around the (3 3 3) Bragg peak at a temperature of $T = 20$~K. The inset shows a section of reciprocal space around the (3,3,3) peak. (c) Relative change in intensity $(\Delta I(t)/I_0)$ for representative wavevectors along the [0 0 1] direction indicated by the color coded dots in (b). The wavevectors are labeled next to each trace.
}
\end{figure*}

Here we probe the mesoscopic polar dynamics of \sto{} at nanometer lengthscales and femtosecond timescales using ultrafast hard x-ray diffraction and diffuse scattering \cite{trigoFouriertransformInelasticXray2013} at the Linac Coherent Light Source (LCLS)\cite{Emma2010First}. 
To induce the transient states\cite{liTerahertzFieldInduced2019} we use single-cycle pulses of THz radiation resonant with the soft modes of \sto{} in the quantum paraelectric regime, as illustrated in Fig \ref{fig:1}a (see Methods for details).
To characterize the polar properties of the transient lattice distortions we devised a scheme to switch reproducibly between opposite polarities of the generated THz field. Related methods have demonstrated the ability to manipulate ferroic orders \cite{kubacka2014large}. In this work we show that when combined with diffuse x-ray scattering, this method is an incisive probe of polar modes and inversion symmetry (see Fig \ref{fig:1}a) breaking at the nanoscale.

In Fig. \ref{fig:1}b we show the equilibrium x-ray intensity around the $(3,3,3)$ Bragg peak along the $[0,0,1]$ direction (pseudo cubic notation), at a temperature of $T = 20$~K. Below $T = 105$~K, \sto adopts a tetragonal structure whose twin domains can be identified in the scattering \cite{lytle1964x,haywardCubictetragonalPhaseTransition1999} (see supplementary information). Here we focus on domains that have their c-axis oriented in the direction of the sample normal, $[0,0,1]$. 
The reciprocal space map in the inset of Fig. \ref{fig:1}b shows that the intensity is characterized by strong directional scattering along the sample normal, which represents heterogeneity in the lattice mostly in the $[0,0,1]$ direction.

Fig. \ref{fig:1}c shows the differential x-ray intensity $\Delta I(t)/I_0$ as a function of pump-probe delay at different representative wavevectors along the $[0,0,1]$ direction, where $\Delta I(t)=I(t)-I_0$ and $I_0$ is the x-ray intensity before the THz excitation. 
Most of the signal is localized in the range $\xi = 0.03 - 0.06$~reciprocal lattice units (rlu), corresponding to lengthscales of $10$-$20$~nm, while the $\xi = 0$ signal at the Bragg peak (gray line) does not show an appreciable dynamic response. 
This is surprising since the strongest resonance to the spatially uniform incident field is expected to be from the transverse optical (TO) mode at $\xi = 0$\cite{liTerahertzFieldInduced2019}. The lack of signal at long wavelengths (gray line) means that we do not observe a strong uniform response of the system as a whole. Furthermore, momentum conservation forbids direct THz excitation at high wavevectors in a homogeneous sample. Thus these signals arise from inhomogeneous excitation by the pump, which can only occur through pre-existing inhomogeneities in the sample.
The $10$-$20$~nm lengthscale is not determined by the attenuation length of the THz radiation, which is on the order of $ 500$~nm, but as we discuss below, coincides with wavevectors at which the TA and TO modes are most strongly coupled\cite{fauqueMesoscopicFluctuatingDomains2022,stirlingNeutronInelasticScattering1972,vacherAnomaliesTAPhononBranches1992}.

\begin{figure}
\centering
\includegraphics[width=0.9\linewidth]{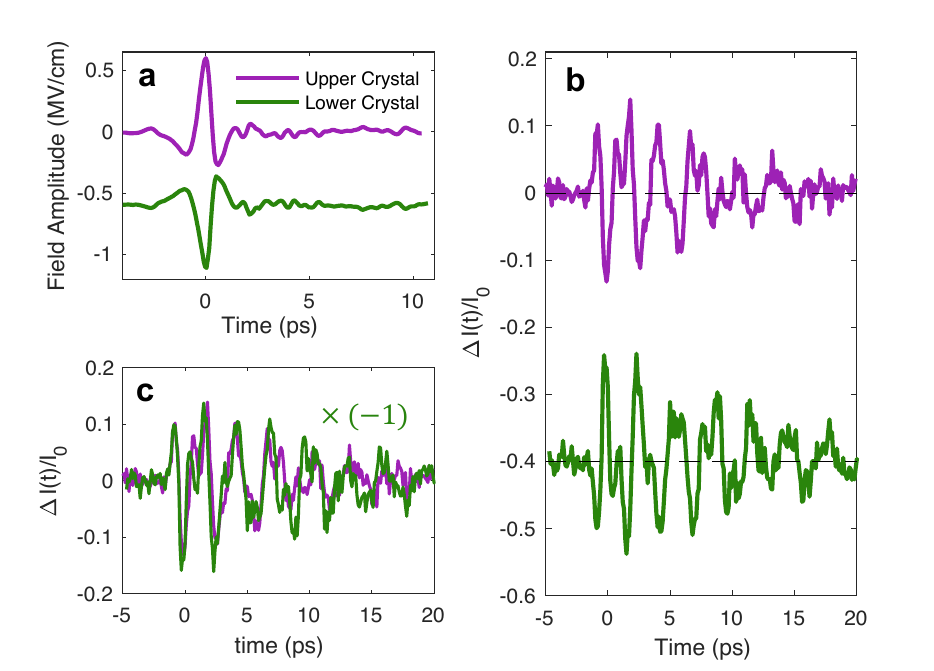}
\caption{\label{fig:2}
{\bf Odd symmetry of the ultrafast response.} (a) Electro-optic sampling measurement of the THz pulse profile. Purple (green) trace represents the THz pulse generated by the upper (lower) \lbo{} crystal. The effect of internal reflection inside the EOS crystal were removed by a standard procedure\cite{naftalyMethodRemovingEtalon2007}.
(b) Dynamics of $\Delta I/I_0$ following THz excitation, integrated between 0.03 and 0.04 rlu, where the purple and green traces represent the response following the THz generated by the upper and lower \lbo{} crystals, respectively.
(c) Same as (b) but with the signal for the lower crystal inverted.
}
\end{figure}

To characterize these THz-induced excitations, we devised a scheme to invert the polarity of the generated THz field by switching between two \lbo{} prisms, mounted with their ferroelectric polarizations pointing in opposite directions as shown in  Fig. \ref{fig:1}a. 
Fig \ref{fig:2}a shows the electro-optic sampling traces of the two THz pulses generated by this setup, which are inverted and otherwise nearly identical. 
In Fig. \ref{fig:2}b we show $\Delta I(t)/I_0$ following excitation by the two THz waveforms, integrated over wavevectors $\xi = 0.03 - 0.04$ rlu, using the same colors as Fig. \ref{fig:2}a to reference the corresponding field polarity.
For better comparison, Fig. \ref{fig:2}c presents the same data with one of the traces flipped, which shows that $\Delta I(t)$ inverts when the THz field is inverted, directly revealing that these oscillations originate from polar collective modes. 
These polar modes break both local inversion and translational symmetry (conservation of the crystal momentum) as they evolve on their natural timescales, and enable direct dipole coupling to the long-wavelength THz radiation.
We associate this inhomogeneity with polar nanoregions which play an important role in the quantum paraelectric phase of \sto{} \cite{Bussmann2009precursor,bussmann2006intrinsic,Laguta2005nmr}.

\begin{figure}
\centering
\includegraphics[width=0.6\linewidth]{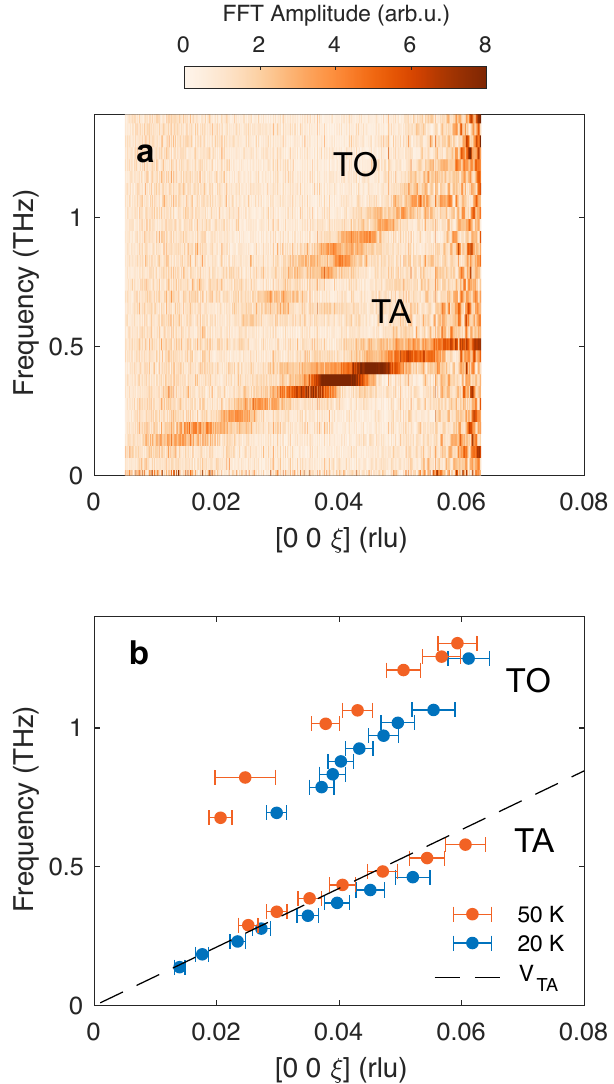}
\caption{\label{fig:3}
{\bf Dispersion relation of coherent excitations.} (a) Colormap representing the magnitude of the Fourier transform of time domain data, similar to those presented in Fig. \ref{fig:1}c, at different wavevectors. (b) Dispersion relation obtained by fitting the peak position for each frequency bin in (a). The error bar in wavevector include statistical and estimated systematic errors due to possible sample misalignment.
The dashed line represents the speed of sound of TA modes from Brillouin scattering data\cite{hehlenHighfrequencyElasticConstants1996}. 
}
\end{figure}

Further insight into the nature of these excitations can be gained by identifying how the different frequency components of the oscillating signals disperse with wavevector. The color map in Fig. \ref{fig:3}a represents the absolute value of the Fourier transform of $\Delta I(t)/I_0$ for different wavevectors along the $[0,0,1]$ direction. We observe two dispersive modes around 0.4 and 1 THz, which we assign to the transverse acoustic (TA) and transverse optical (TO) branches, respectively \cite{stirlingNeutronInelasticScattering1972,vacherAnomaliesTAPhononBranches1992,fauqueMesoscopicFluctuatingDomains2022}.
The spectral intensity is strongest in the TA branch, which plays a prominent role in the polar response reported in Figs. \ref{fig:2}b-c. In Fig. \ref{fig:3}b we show the dispersion relation at 20 K (blue) and 50 K (red) obtained from our measurements (Fig. \ref{fig:3}a). The black dashed line shows the linear dispersion of the TA branch measured by Brillouin scattering \cite{hehlenHighfrequencyElasticConstants1996} extrapolated to high wavevectors.
The TO branch in Fig. \ref{fig:3}b shows softening with decreasing temperature characteristic of the incipient FE\cite{fauqueMesoscopicFluctuatingDomains2022,shiraneLatticeDynamicalStudy1101969}. Notably, the acoustic dispersion deviates from the extrapolated speed of sound at $\xi > 0.03$~rlu at 20~K. A similar anomaly has been observed by neutron scattering in equilibrium\cite{stirlingNeutronInelasticScattering1972,vacherAnomaliesTAPhononBranches1992,fauqueMesoscopicFluctuatingDomains2022} and was associated with some form of coupling between the TO and TA branches. Comparing with Fig \ref{fig:3}a, it is clear that $\xi > 0.03$~rlu is also where the TA branch exhibits the strongest response to the pump and hence the strongest polar character.
These results suggest a common physical origin for the frequency softening of the acoustic modes and their strong polar character observed here.

\begin{figure*}
\centering
\includegraphics[width=0.9\textwidth]{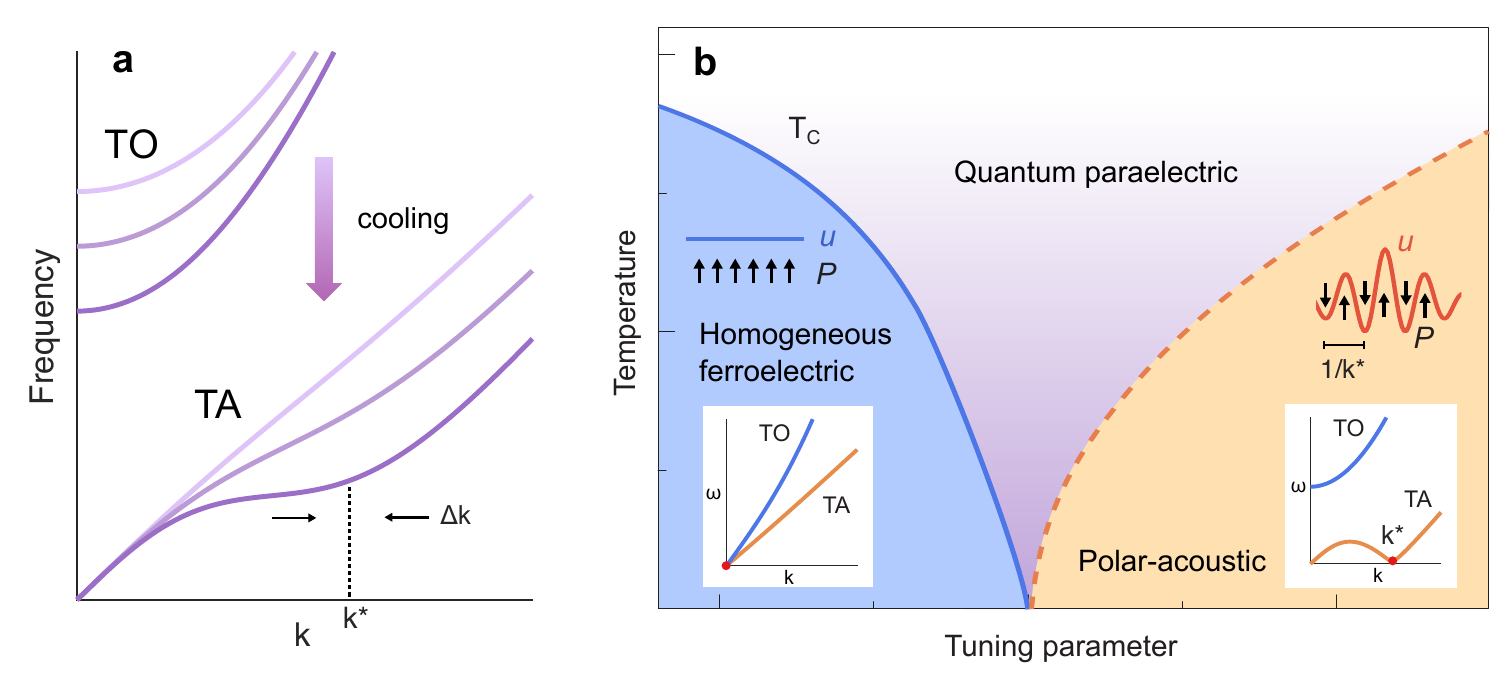}
\caption{\label{fig:4} {\bf Lattice instability towards a modulated structure.}
(a) Schematic illustration of the dispersion of transverse phonon modes. As temperature decreases the TO branch softens and the TA develops a kink at finite wavevector $k^*$, indicative of an instability. These excitations are polar and modulate the polarization spatially with a wavelength $\approx 2\pi/k^*$.
(b) Schematic phase diagram of \sto showing the nearby phases around the quantum paraelectric regime. The blue and orange represent the ferroelectric phase and a polar-acoustic regime\cite{guzman-verriLamellarFluctuationsMelt2023,coakQuantumCritical2020}. In the former the polarization is homogeneous while in the latter the strain and polarization are modulated spatially with period $2 \pi/k^*$.  The flexoelectric coupling strength could be taken as the tuning parameter.
}
\end{figure*}

Our measurements offer a deeper understanding of the interplay between strain and polarization in \sto. 
In particular these results provide a robust characterization of their coupling and enable building a new map of the precursor fluctuations of incipient orders as presented in Fig. \ref{fig:4}.
The fact that both the TA and TO signals invert with field polarity and that the signal is linear with the field strength (see supplementary information) imposes strong constraints on the possible form of strain-polarization coupling. 
For example, electrostrictive coupling is quadratic in the ferroelectric polarization\cite{xuStraininducedRoomtemperatureFerroelectricity2020} and would give rise to nonlinearity and harmonics in the oscillatory signals and to a non-inverting response, none of which are observed here.
Instead, our observations indicate that the coupling must be bilinear in strain and polarization. In a paraelectric system, symmetry constrains this coupling to be flexoelectric, that is between the ferroelectric polarization and the gradients of the strain \cite{tagantsevFlexoelectricitySolidsTheory2016,zubkoFlexoelectricEffectSolids2013,yudinFundamentalsFlexoelectricitySolids2013,wangFlexoelectricitySolidsProgress2019,Biancoli2015}. 
Flexoelectricity explains our main observations: (1) the odd symmetry of both modes upon inversion, (2) the linearity of both TA and TO with field strength, and (3) the softening to the acoustic branch at finite wavevectors.

We summarize these results qualitatively in Fig \ref{fig:4}a, showing an illustration of the dispersion relation at different temperatures, taking into account the flexoelectric coupling\cite{tagantsevFlexoelectricitySolidsTheory2016,zubkoFlexoelectricEffectSolids2013,guzman-verriLamellarFluctuationsMelt2023}. The TO branch softens considerably at zone center upon cooling\cite{yamadaNeutronScatteringNature1969}, and corresponds to the order parameter of the incipient ferroelectric phase. 
Simultaneously, coupling between the TO and TA branches causes the softening of the TA branch around the wavevector $k^*$, and an increase of its polar character. 

In Fig \ref{fig:4}b we show a schematic phase diagram of \sto\cite{rowleyFerroelectricQuantumCriticality2014,coakQuantumCritical2020,mullerIndicationNovelPhase1991,guzman-verriLamellarFluctuationsMelt2023} as a function of temperature and the flexoelectric coefficient that governs the TO-TA coupling strength. 
The diagram illustrates that the quantum paraelectric regime separates a homogeneous FE phase, where the zone-center TO mode softens to zero frequency, from a putative polar-acoustic regime\cite{coakQuantumCritical2020} where the acoustic branch softens at $k^*$ and enhances strongly the polarization and strain modulation at this wavevector. While the soft TO mode is a precursor of the spatially homogeneous ferroelectric phase (light blue in Fig. \ref{fig:4}b), the polar-acoustic modes observed here are the precursors of a possible modulated polar-acoustic phase (orange in Fig. \ref{fig:4}b)\cite{guzman-verriLamellarFluctuationsMelt2023,coakQuantumCritical2020}. 

Our measurements unveil the local polar excitations in \sto which shape its quantum critical behavior and can foster unconventional superconductivity\cite{Hameed2022}. These polarization density waves, which arise from a flexoelectric interaction, provide direct evidence of a modulated polar instability in \sto{} which may suppress ferroelectricity\cite{guzman-verriLamellarFluctuationsMelt2023}. 
More generally, nanoscale instabilities often herald exotic and sought-after phases in materials with complex phase diagrams.  Our results show that time-resolved measurements with access to finite wavevectors can reveal new instabilities that are key to unlocking and controlling novel phases of complex materials.

\newpage

\section*{Methods}
\subsection*{Experimental details}
Single crystal \sto{} was commercially obtained from MTI Corp. The $10\times10\times0.5$ mm$^3$ sample was Verneuil-grown, with two sided epi-polished (001) surfaces. Measurements were taken at sample temperatures of 20 K and 50K in the cryogenic compatible vacuum chamber at the x-ray pump probe (XPP) end station at Linac Coherent Light Source (LCLS) \cite{cholletXrayPumpProbe2015}. The sample was mounted inside a vacuum compatible chamber with translation and rotation motions required to reach the desired x-ray diffraction and scattering conditions. A dedicated copper mount was fabricated for this purpose, pre-tilting the sample by $10^\circ$, allowing us to reach the (3,3,3) diffraction peak at 10.4 keV and measure the scattering around it. The crystal was cooled by a helium flow cryostat. The cold finger was connected with flexible copper braids to the dedicated mount and the temperature was measured by a diode fixed close to the sample position.

Intense THz fields were generated by optical rectification of 120 fs, near IR pulses centered at 800 nm in \lbo{} using the tilted-pulse-front technique\cite{heblingGenerationHighpowerTerahertz2008,fulopGenerationMJlevelUltrashort2011,yehGeneration10mJUltrashort2007,hoffmannIntenseUltrashortTerahertz2011}. To generate THz fields with inverted polarities two \lbo{} prisms were mounted vertically one on top of the other outside the vacuum chamber, with their FE polarizations pointing in opposite directions. A motorized vertical translation stage was used to switch between the prisms in the 800 nm beam-path. The temporal profile of the two inverted fields at the sample position was measured using electro-optic sampling in a 100 μm GaP crystal (Fig. \ref{fig:2}a). Intensity and polarization of the THz pulses were controlled by a pair of wire-grid polarizers. An off-axis parabolic mirror placed inside the chamber was used to focus the pulses into a $\sim1$ mm diameter spot on the sample position. 

The lattice dynamics was probed by diffraction and scattering of 50 fs, 10.4 keV hard x ray pulses focused to a 300 $\mu$m diameter spot. A small hole in the THz focusing mirror for transporting the x-rays allowed colinear propagation of the x-ray and THz pulses, which impinged on the sample at an angle of $\simapprox{}30^\circ$. The scattered photons were recorded through a Kapton window by a Jungfrau area detector positioned outside the chamber 300 mm from the sample. To match the pumped and probed volumes, while maximizing the pump fluence, we implemented a grazing exit geometry. The exit angle was set to $\simapprox{}1^\circ$ with respect to the sample surface such that the detected x ray photons originate from a region $< 1 ~\mu$m into the sample, comparable to the penetration depth of the THz pulse in \sto{} at low temperatures.

\section*{Acknowledgments}
We acknowledge insightful discussions with Beno\^{i}t Fauqu\'{e}, Gian Guzm\'{a}n-Verri and Peter Littlewood.
The experimental work was supported by the U.S. Department of Energy, Office of Science, Office of Basic Energy Sciences through the Division of Materials Sciences and Engineering through Contract No. DE-AC02-76SF00515 (G. O., V. K., Y. H., G.dP., R. D., D. A. R. and M. T.) and Contract No. DE-SC0019126 (Z. Z., R. C. and K. A. N.). 
Work at Brookhaven is supported by the Office of Basic Energy Sciences, Materials Sciences and Engineering Division, U.S. Department of Energy (DOE) under Contract No. DE-SC0012704.
Use of the LCLS was supported by the U.S. Department of Energy, Office of Science, Office of Basic Energy Sciences under Contract No. DE-AC02-76SF00515. 
S. L. J. acknowledges support from the Swiss National Science Foundation under project grant 200020\_192337. Work at Harvard University was supported by the U.S. Department of Energy, Office of Basic Energy Sciences, Early Career Award Program, under Award No. DE-SC0022883.
J.S. acknowledges supported by the NSF under REU supplement 2133686 under award 2019014.

\newpage
\begin{singlespace}
\title{Supplementary}

\maketitle
\end{singlespace}

\pagebreak

\section*{Identifying Twin Domains}
Upon cooling below the structural phase transition at $T=105K$, an antiferrodistortive (AFD) instability induces a low-symmetry tetragonal phase in \sto \cite{lytle1964x,haywardCubictetragonalPhaseTransition1999}. The distortion can develop along any of the three equivalent cubic crystallographic directions, leading to the formation of three potential twin domains. As a result, the Bragg reflections separate into three distinct peaks, originating from the different domains illuminated by our x-ray probe. In Fig. \ref{twin_domains}a we show the detector image of the x-ray scattering, with the sample oriented close to the nominal (3,3,3) reflection. Two distinct features appear in the image, marked by a red square and a green square. In Fig. \ref{twin_domains}b we present the integrated intensity of these features versus the angle of one of the sample rotation motors exhibiting three distinct peaks corresponding to the three different orientational domains. To identify the orientation of the twin responsible for each peak we use the measured \sto lattice constants \cite{kiatRietveldAnalysisStrontium1996}, distinguishing between the three twins by directing the long axis along the three crystallographic directions and calculating the (3,3,3) Bragg reflection for each orientation. We compare the diffraction direction and sample orientation of the calculated reflections to those of the experimental peaks in Fig. \ref{twin_domains}b. Peak 1 is identified with a domain where the long axis is oriented perpendicular to the surface normal of the sample and parallel to the THz polarization, peak 2 with a domain where the long axis is oriented perpendicular to the surface normal and perpendicular to the THz polarization and peak 3 with a domain where the long axis is parallel to the surface normal. In the main text we focus on twin 3, which has the strongest signal.

\renewcommand{\thefigure}{\textbf{S\arabic{figure}}}
\setcounter{figure}{0}

\begin{figure*}[h] 
\centering
\includegraphics[width=0.8\linewidth]{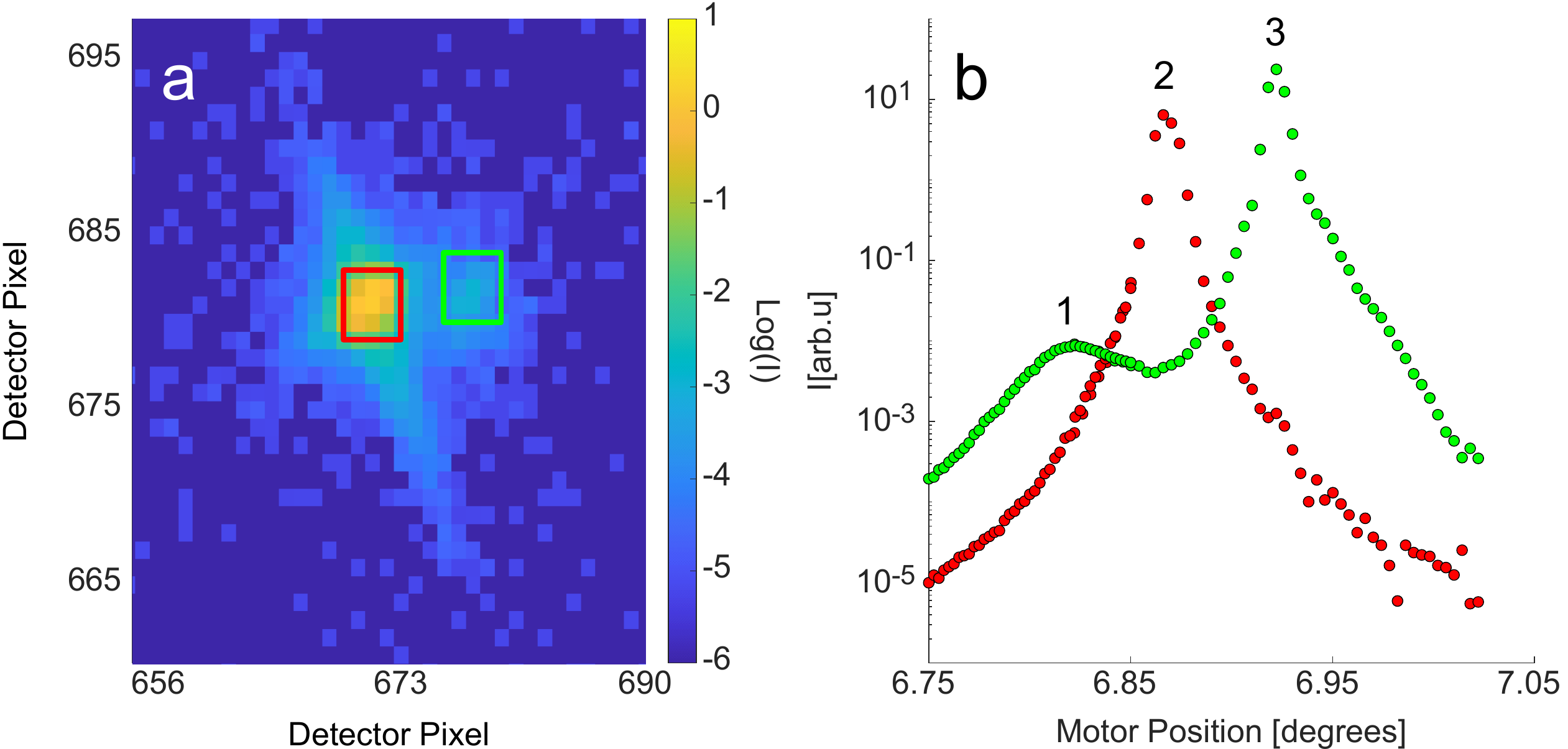}
\caption
{\label{twin_domains}{\bf Twin domains in the angle scan of the (3,3,3) Bragg peak.} (a) Part of the detector image showing x-ray scattering close to the (3,3,3) Bragg condition. The red and green squares indicate the integration regions for the plots in (b). The sample orientation for this image corresponds to peak 2 in (b). (b) The red and green circles are the integrated intensity in the red and green squares shown in (a) respectively. The intensity is plotted versus the angle of one of the rotation motors which control the sample orientation. 
}
\end{figure*}

\clearpage
\section*{Fluence Dependence}
\begin{figure*}[ht]
\centering
\includegraphics[width=0.8\linewidth]{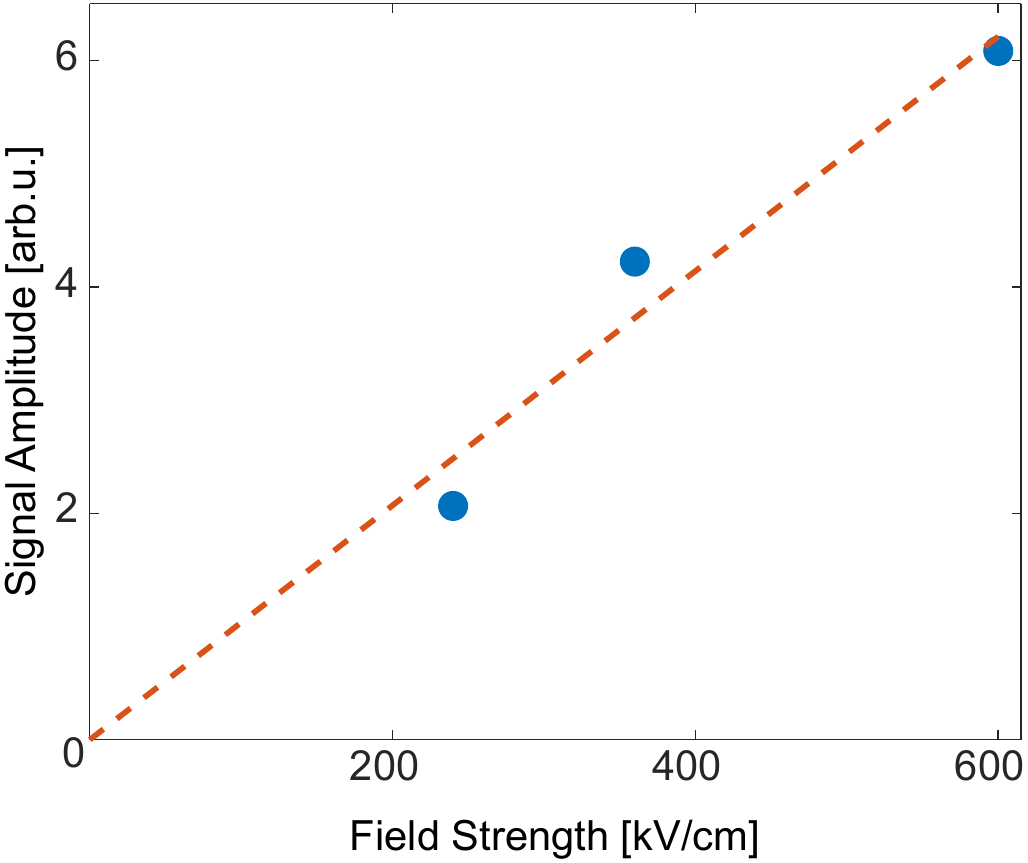}
\caption
{{\bf Fluence dependence of the high wavevector signal.} The blue circles represent the amplitude of the signal versus THz field strength. The broken red line is a linear fit of the data with a zero intercept.
}
\end{figure*}

\end{document}